# Turning non-magnetic two-dimensional molybdenum disulfide into room temperature magnets by the synergistic effect of strain engineering and charge injection

Jing Wu,[ac] Ruyi Guo,[a] Daoxiong Wu,[b]* Xiuling Li,[ad]* and Xiaojun Wu[e]

[a] School of Physics and Technology, Nanjing Normal University, Nanjing 210023, China

[b] School of Marine Science and Engineering, Hainan Provincial Key Lab of Fine Chemistry, School of Chemical Engineering and Technology, Hainan University, Haikou, 570228, China

[c] Yancheng Kangju Road Junior Middle School, Yancheng, 224000, China

[d] National Laboratory of Solid State Microstructures, Nanjing University, Nanjing 210093, China

[e] Hefei National Research Center for Physical Sciences at the Microscale, Collaborative Innovation Center of Chemistry for Energy Materials, CAS Key Laboratory of Materials for Energy Conversion, CAS Center for Excellence in Nanoscience, and School of Chemistry and Materials Sciences, University of Science and Technology of China, Hefei, Anhui 230026, China

## Abstract

The development of two-dimensional (2D) room temperature magnets is of great significance to the practical application of spintronic devices. However, the number of synthesized intrinsic 2D magnets is limited and the performances of them are not satisfactory, e.g. typically with low Curie temperature and poor environmental stability. Magnetic modulation based on developed 2D materials, especially non-magnetic 2D materials, can bring us new breakthroughs. Herein, we report room temperature ferromagnetism in halogenated $MoS_2$ monolayer under the synergistic effect of strain engineering and charge injection, and the combined implementation of these two processes is based on the halogenation of $MoS_2$. The adsorbed halogen atoms X (X = F, Cl, and Br) on the surface leads to lattice superstretching and hole injection, resulting in $MoS_2$ monolayer exhibiting half-metallic properties, with one spin channel being gapless in the band structure. The Curie temperature of halogenated $MoS_2$ monolayer is 513~615 K, which is much higher than the room temperature. In addition, large magnetic anisotropy energy and good environmental stability make halogenated $MoS_2$ display great advantages in practical spintronic nanodevices.

# Introduction

Spintronics, exploiting the spin degree of electrons as the information vector, is an ideal field for future information technology.[1] Two-dimensional (2D) materials can provide an excellent platform for spintronics research due to their distinctive spin-dependent properties.[2,3] For example, weak spin-orbit coupling enables enhanced spin diffusion length and lifetime in graphene channels, opening up new prospects for long-distance spin communication.[4] The room-temperature spin hall effect can be achieved in graphene/$MoS_2$ van der Waals heterostructures, enabling efficient spin-charge conversion and spin-polarized current control.[5] The inversion symmetry breaking together with spin-orbit coupling leads to the coupling of spin and valley in monolayers of group-VI dichalcogenides, providing a route towards the integration of valleytronics and spintronics.[6] Two-dimensional materials thus open up exciting opportunities for a variety of future spintronic applications. However, quite a few challenges also exist in this field. One of the most prominent problems is the search for an ideal two-dimensional magnetic material that operates at room temperature.[7]

Several two-dimensional magnetic materials have been successfully synthesized experimentally. Typical examples are the $CrI_3$ monolayer and $Cr_2Ge_2Te_6$ bilayer observed by magneto-optical Kerr technique in 2017, which opened the era for the research of intrinsic 2D ferromagnetic materials.[8,9] Subsequently, $Fe_3GeTe_2$, $VSe_2$, $CrSe_2$ and other 2D magnetic films were reported successively in the experiment, which continuously enriched the library of 2D magnetic materials.[10-12] However, the number of such intrinsic 2D magnetic materials is still insufficient. At the same time, the critical temperature of these 2D magnetic materials is usually well below room temperature, which also limits their practical applications. The Curie temperature of $CrI_3$ monolayer is 45 K and that of $Cr_2Ge_2Te_6$ monolayer is 61K.[8,9] Moreover, most of these materials face the difficulty of poor environmental stability. The $CrI_3$ monolayer was observed experimentally to rapidly form small droplets at its edges and gradually degrade in less than 20 seconds.[13,14] To sum up, it is of great significance to develop and design suitable new 2D magnetic materials from realistic considerations.

Magnetic modulation based on existing developed 2D materials can bring us new breakthroughs and offer exciting prospects for future spintronic devices. For example, the heavily electron-doped $Cr_2Ge_2Te_6$ monolayer exhibits ferromagnetic characteristics at temperatures above 200 K, significantly higher than the known Curie temperature of the undoped material, and the magnetic easy axis of this new ground state lies within the layer plane in sharp contrast to the magnetic axis of the undoped $Cr_2Ge_2Te_6$ monolayer pointing out of the plane.[15] These findings highlight the potential of electrostatic gating in magnetic modulation of 2D materials. Similar results were obtained in $Fe_3GeTe_2$ thin flakes as well. In addition to electrostatic gating,[15,16] defects,[17,18] strain engineering,[19,20] interlayer stacking[21,22] *et al.* can also be applied to magnetic modulation in 2D materials. It can be proved that the coercive field, Curie temperature, and transition temperature between single- and labyrinthine- domain states of $Fe_3GeTe_2$ thin flakes increase remarkably under the tensile strain.[20] The strain-dependent ultrasensitive magnetization promotes the development of new straintronic device applications. The strain-induced switching of magnetic states also can be found in $VX_2$, $NbX_2$ (X = S, Se) and other 2D magnetic materials.[19,23]

However, the synergies of different operations are often overlooked. For instance, the surface functionalization of 2D materials, or the influence of substrates and the interface effects brought about by stacking often cause not a single aspect of the impact. The system may possess both strain and charge transfer, or defects. Therefore, we hope to better realize the magnetic modulation of 2D materials from the perspective of synergistic effects. In this paper, the chemically modified $MoS_2$ monolayers with halide elements X (X = F, Cl and Br) have been theoretically investigated. The results show that the chemisorption of halogen atoms on the surface of $MoS_2$ can lead to the structural superstretching and heavy hole injection, which transform $MoS_2$ monolayers from non-magnetic to room-temperature ferromagnetic with half semi-metallic band structure. High Curie temperature ($T_C$), large magnetic anisotropy energy and good environmental stability make halogenated $MoS_2$ show great advantages in practical spintronic nanodevices. The findings also illustrates prospect of realizing high-performance spintronic devices based on non-magnetic two-dimensional materials.

**Calculation details**

The first-principles calculations are performed by a spin-polarized density-functional theory (DFT) method using the generalized gradient approximation (GGA) proposed by Perdew-Burke-Ernzerhof for the exchange and correlation energy, as implemented in the Vienna Ab initio Simulation Package (VASP).[24,25] The projected augmented wave (PAW) method is used with an energy cutoff of 520 eV.[26] The vacuum area perpendicular to the surface is set to at least 15 Å to avoid interaction between two adjacent layers. To well describe the strongly correlated d electrons of Mo atoms, the Hubbard U correction (DFT+U) and the screened hybrid HSE06 functional were employed in structural relaxation and the calculation of electronic properties, respectively.[27,28] The van der Waals corrections by the Grimme's DFT-D3 method was also included.[29] All atomic coordinates are fully relaxed until the total energy is converged to $10^{-6}$ eV and the atomic forces is converged to 0.01 eV/Å. The first Brillouin zone is sampled with a Γ-centered k-point mesh of 12 × 12 × 1. Born-Oppenheimer molecular dynamics (BOMD) simulations were performed in the canonical ensemble using a Nosé thermostat method with the temperature of 300 K.[30] The time step is 1 fs and the total simulation time is 5 ps. The phonon dispersion spectrum was calculated by using the finite displacement method on a 6 × 6 × 1 supercell within the PHONOPY code.[31] Heisenberg model-based Monte Carlo (MC) simulations were lastly used to estimate the Curie temperatures of chemically modified $MoS_2$ monolayers.[32,33]

**Results and discussion**

$MoS_2$ is one of the most studied transition metal dichalcogenides. Several prototypes of $MoS_2$ layers have been reported, of which 2H and 1T phases have received a lot of attention in theoretical calculations and experiments.[34-36] Both phases are non-magnetic, where 2H-$MoS_2$ is a semiconductor and 1T-$MoS_2$ is metallic. Here, the fluorinated $MoS_2$ ($MoS_2F_2$) was first discussed in detail as a representative. The interesting results show that the most stable structural phase of $MoS_2F_2$ is the 1T phase (see Figure 1a) rather than the 2H phase (see Figure S1a), which is inconsistent with the case of primary MoS$_2$. The energy of 1T-phase MoS$_2$F$_2$ (1T-MoS$_2$F$_2$) is about 0.131 eV lower than that of 2H phase MoS$_2$F$_2$ (2H-MoS$_2$F$_2$). Figure 1b and c display the calculated phonon spectrum of 1T-MoS$_2$F$_2$ and the BOMD simulation. The phonon spectrum clearly has no imaginary frequency, and there is no obvious structural deformation when the BOMD simulation temperature is set to 300 K. These simulations forcefully prove that the 1T-MoS$_2$F$_2$ has good thermodynamic stability. However, the phonon spectrum of the 2H-MoS$_2$F$_2$ has a large imaginary frequency (see Figure S1b). Therefore, in the following study, we focus on the 1T phase halogenated MoS$_2$ monolayers.

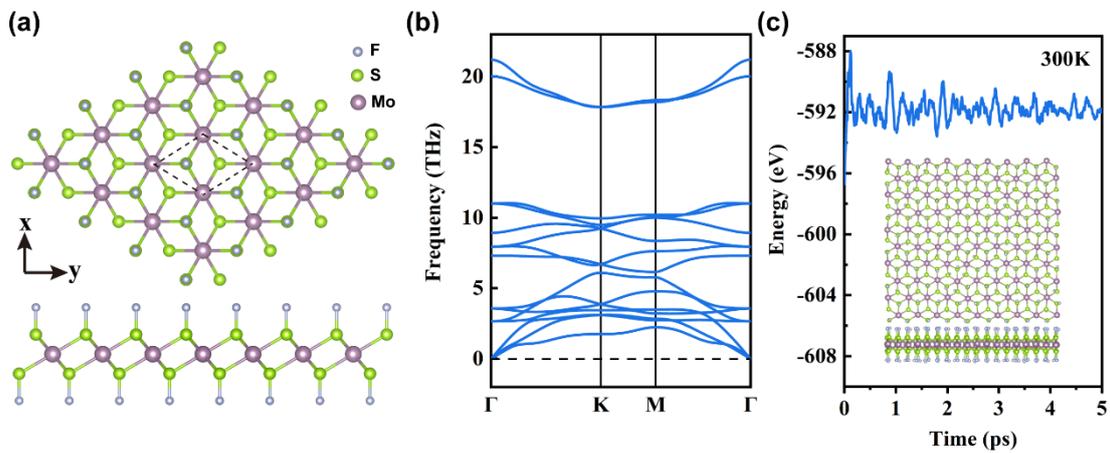

**Figure 1.** (a) The top and side views of fluorinated MoS$_2$ monolayer with T phase (1T-MoS$_2$F$_2$). (b) The phonon spectrum of 1T-MoS$_2$F$_2$ monolayer. (c) The BOMD simulation of 1T-MoS$_2$F$_2$ monolayer at the temperature of 300 K. The blue curve monitors the change in energy. The insert displays the structure of MoS$_2$F$_2$ after 5ps.

In addition to thermodynamic stability, the obvious changes of the lattice of halogenated MoS$_2$ deserve our attention. The optimized lattice constants of the original 1T-MoS$_2$ are a = b = 3.163 Å, while the fluorinated T-phase MoS$_2$ has a large lattice stretching, where the lattice values can reach a = b = 3.654 Å. The calculated bond length and bond angle of Mo-S-Mo also change after surface fluorination. The band angle of Mo-S-Mo in the primary 1T-MoS$_2$ is 81.2°, while the band angle of Mo-S-Mo in the 1T-MoS$_2$F$_2$ is 95.9°. The overstretching effect caused by this surface engineering can affect the crystal field, resulting in changes in the splitting of Mo's d electrons, which give rise to interesting electronic and magnetic properties.

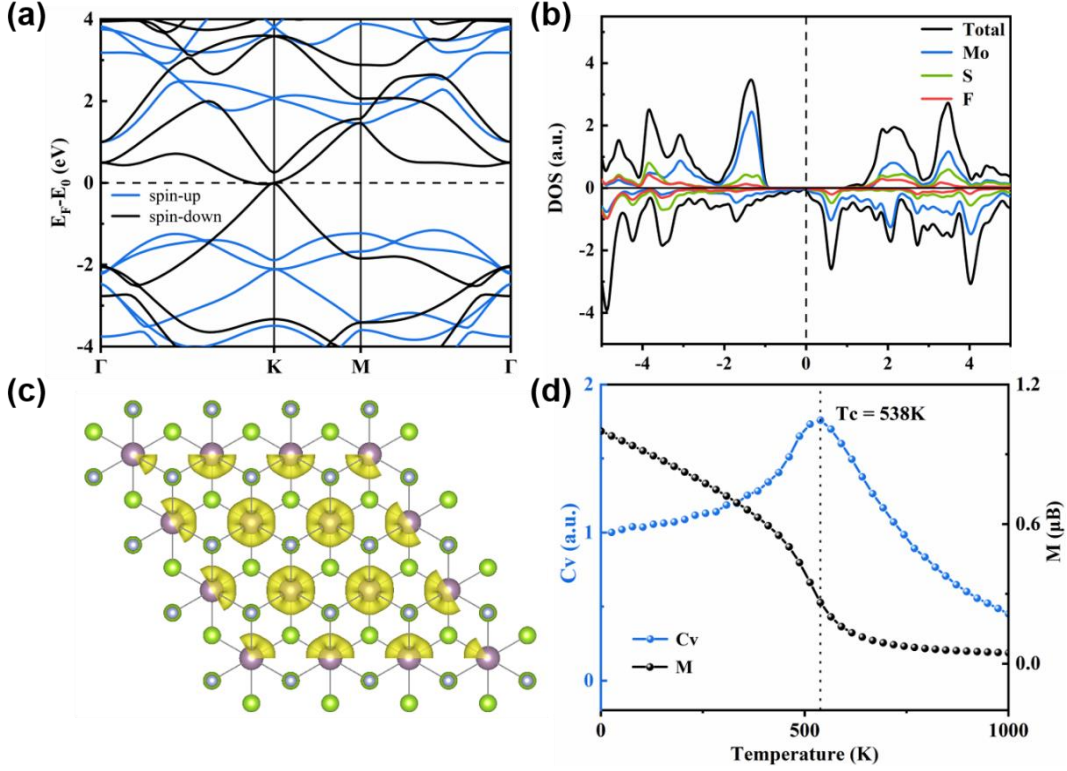

**Figure 2.** (a) Spin-polarized band structure of 1T-MoS$_2$F$_2$ monolayer calculated using the HSE06 function with the Fermi energy level set to 0 eV, indicated by the black dashed line. (b) Total density of states of 1T-MoS$_2$F$_2$ monolayer and the partial density of states of each atom. (c) Spin charge density distribution of 1T-MoS$_2$F$_2$ monolayer. (d) The simulated magnetic moment (M) and specific heat (C$_v$) with respect to temperature for 1T-MoS$_2$F$_2$ monolayer.

Figure 2 displays the simulated electronic properties and magnetic states of 1T-MoS$_2$F$_2$ monolayer. The spin-polarized band structure and the calculated density of states show that 1T-MoS$_2$F$_2$ is ferromagnetic with half-metallic band structure, where one spin channel is gapless and the other has a wide bandgap of 2.15 eV. The spin-polarization ratio near the Fermi energy level is about 100%. Figure 2b shows the density of states (DOS) of 1T-MoS$_2$F$_2$. Both the partial density of states (PDOS) of each atom and the calculated magnetic moment show that 1T-MoS$_2$F$_2$ is magnetic, and the origin of the magnetism is mainly contributed by Mo atom. The magnetic moment of each cell is about 2 $\mu_B$, and the local magnetic moment of the Mo atom is about 1.92

$\mu_B$. By contrast, the primary MoS$_2$ is non-magnetic regardless of whether it is in the H phase or T phase. In order to determine the magnetic ground state of 1T-MoS$_2$F$_2$, two magnetic orders, that is, ferromagnetic (FM) and antiferromagnetic (AFM) coupling, are considered in a $\sqrt{3} \times 1 \times 1$ supercell, as seen in Figure S3. The results show that the 1T-MoS$_2$F$_2$ monolayer has a FM ground state, which is more stable than AFM state of 166 meV per unit cell. The 1T-MoS$_2$F$_2$ monolayer is half-metallic, giving rise to particular advantages in spintronics applications. The gapless band structure of one spin channel may exhibit interesting topological behaviors, which also enriches the application of 1T-MoS$_2$F$_2$ in the field of low-dimensional spintronic nanodevices.

The magnetic anisotropy energy (MAE) is an important parameter of magnetic materials to determine the low-temperature magnetic orientation, which is directly related to the thermal stability of magnetic data storage.[37] Calculations including spin−orbital coupling are therefore performed to estimate the easy magnetization axis and the MAE of 2D 1T-MoS$_2$F$_2$ monolayer. As a result, the 2D 1T-MoS$_2$F$_2$ crystal have in-plane magnetic anisotropy with the easy magnetization axis parallel to the $x$ direction (100) and the MAE is about 804 $\mu$eV/Mo, which is comparable to that of the CrI$_3$ monolayer (980 $\mu$eV/Cr) and larger than those of CrCl$_3$ and CrBr$_3$ monolayers (25 $\mu$eV/Cr, 160 $\mu$eV/Cr).[38,39] The sizable MAE indicates that the 1T-MoS$_2$F$_2$ nanosheet is suitable for magnetoelectronics applications.

T$_C$ is another key parameter for the practical application of spintronic devices. The Monte Carlo (MC) simulation based on the Heisenberg model is adopted to estimate the T$_C$ of 1T-MoS$_2$F$_2$, which has been used in previous simulations to reliably predict the magnetic critical temperature of 2D crystals (such as CrX$_3$, X = Cl, Br, I).[32,38] The spin Hamiltonian is defined as $H = \sum_{<i,j>} JS^2$, where $J$ is the exchange parameter between the two neighbor Mo atoms, and $S$ is the spin vector of each Mo atom ($S = 1$). Considering only the exchange of the nearest Mo atoms, the magnetic states of 1T-MoS$_2$F$_2$ can be described as $E_{FM} = E_0 + 6JS^2$ and $E_{AFM} = E_0 - 6JS^2$, respectively, where $E_{FM}$ is the energy of FM state, and $E_{AFM}$ is the energy of AFM state, and $E_0$ is the energy without magnetic coupling. Using the energy difference between FM and

AFM (332 meV per $\sqrt{3} \times 1 \times 1$ supercell), the estimated Heisenberg exchange parameter $J$ is -41.5 meV, which is much higher than that of 2D CrI$_3$ (−2.7 meV),[38] implying an ideal critical temperature. A large supercell of $40 \times 40 \times 1$ with periodic boundary condition is used in the MC simulation. After the system reaches equilibrium at a given temperature, the specific heat capacity $C_v$ can be calculated, as seen in Figure 2d. The $T_C$ of 2D magnet can be estimated from the peak positions of specific heat $C_v$. For 1T-MoS$_2$F$_2$ nanosheet, the $T_C$ is about 538 K, which is significantly higher than that of previously reported 2D intrinsic FM crystals. The $T_C$ above room temperature allows 1T-MoS$_2$F$_2$ working at ambient environment, which possesses great practical promising.

The following analysis will reveal the magnetic mechanism of fluorinated 1T-MoS$_2$. As mentioned above, surface fluorination introduces significant overstretching in 1T-MoS$_2$, which inevitably leads to changes in the coordination field where Mo atoms reside, thereby affecting the arrangement of $d$ orbitals. The occupation of $d$ electrons is closely related to the magnetism of the material. Therefore, the electronic band structures of 1T-MoS$_2$ under different lattice stretching (see Figure S2) and the corresponding projected density of states of Mo atoms contributed by $d$ orbitals (see Figure 3) were calculated. It is obvious that the electronic structure of 1T-MoS$_2$ changes with the lattice stretching. In the absence of lattice stretching or with a small lattice stretching, 1T-MoS$_2$ is metallic and non-magnetic. But when the lattice stretch reaches 4%, 1T-MoS$_2$ begins to exhibit magnetic properties. At this point, the Mo atom has a local magnetic moment of 1.391 $\mu_B$. As the stretching degree increases, the local magnetic moment of Mo atoms increases slightly. At the same time, 1T-MoS$_2$ undergoes a transition from a non-magnetic metal to a magnetic metal, and then transforms into a magnetic semiconductor. We focused on the causes of magnetism, so we chose 4% lattice stretching as a representative to study the arrangement of the $d$ orbitals, which is precisely the critical case where 1T-MoS$_2$ has just demonstrated magnetism.

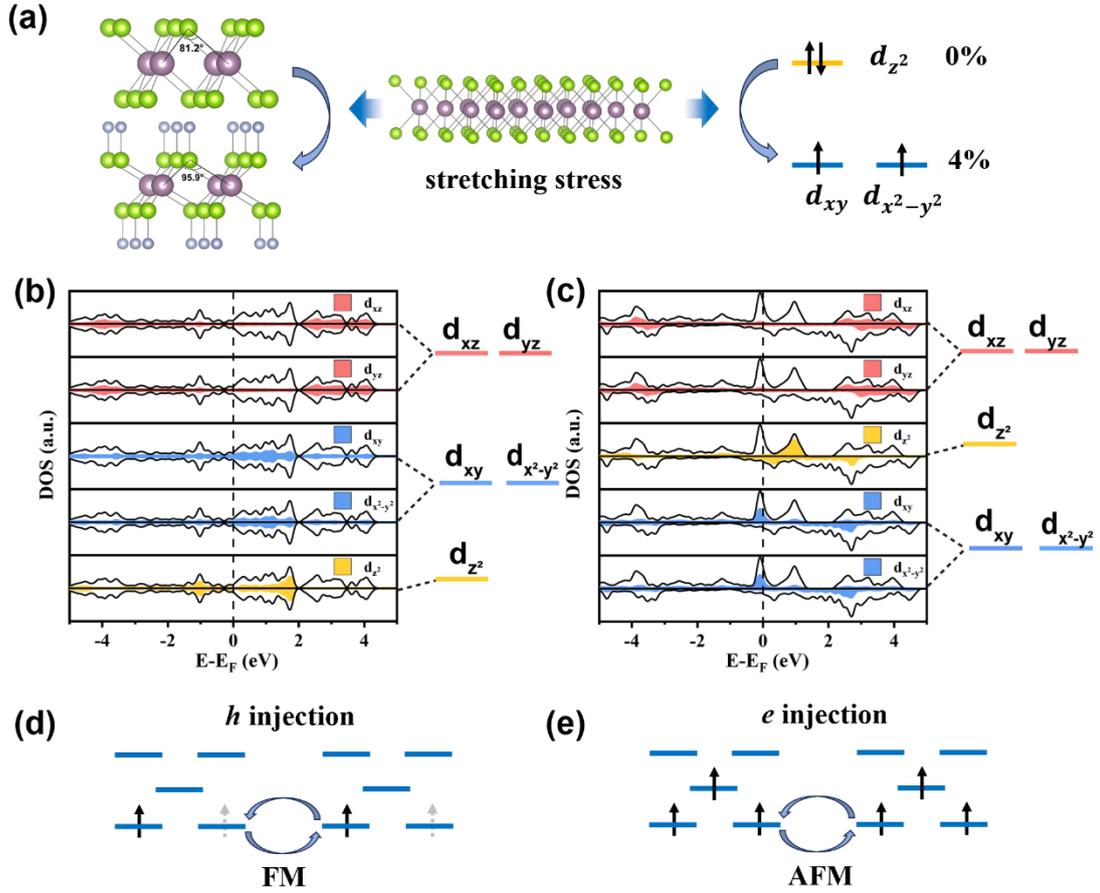

**Figure 3.** (a) Structural models of 1T-MoS$_2$ monolayer with and without fluorination, and the flipping of occupied *d* orbitals of 1T-MoS$_2$ in the case of lattice stretching. (b) The projected density of states of the Mo's d orbitals for 1T-MoS$_2$ without lattice stretching. (c) The projected density of states of the Mo's d orbitals for 1T-MoS$_2$ under 4% lattice stretching. (d) Schematic diagram of the *d* orbitals filling of stretched 1T-MoS$_2$ in the case of hole injection and electron injection.

Figure 3 gives a simple diagram of the *d*-orbital flipping of 1T-MoS$_2$ under lattice itretching. The change of magnetic state in 2D 1T-MoS$_2$ can be understood with the crystal field theory. It is clearly seen that, in a deformed octahedral crystal field, the *d* orbitals of Mo atom are splitted into three groups of $d_{z^2}$, $d_{xy}+d_{x^2-y^2}$ and $d_{xz}+d_{yz}$. Figure 3b shows the arrangement of *d* orbitals in 1T-MoS$_2$ without lattice stretching. The occupied states under the Fermi energy level are mainly contributed by non-degenerate $d_{z^2}$. According to the Pauli exclusion principle, the two valence *d* electrons can only be antiparallel filled in the $d_{z^2}$ band, thus the Mo atom has no net magnetic moment. But when MoS$_2$ undergoes a certain lattice stretching, the crystal field correspondingly changes. Although the splitting of the *d* orbitals falls into three groups, the relative positions of the occupied and unoccupied states near the Fermi surface are flipped. Under 4% lattice stretching, the degenerate bands of $d_{xy}$ and $d_{x^2-y^2}$ move to positions below the Fermi level, while $d_{z^2}$ band moves to positions above the Fermi level. This means that the two valence *d* electrons can be parallely arranged in the $d_{xy}$ and $d_{x^2-y^2}$ bands, so the Mo atom exhibits a net magnetic moment. The system becomes magnetic. Thus, the fluorinated 1T-MoS$_2$ is magnetic mainly due to the lattice stretching brought about by the surface modification. However, the calculated energies of different magnetic states indicate that the stretched system are AFM regardless of whether 1T-MoS$_2$ is stretched at 4% or larger lattice stretches. Further analysis is necessary.

    The chemical adsorption of halogen elements on the surface of 1T-MoS$_2$ not only brings lattice stretching, but also charge transfer. The effect of charge transfer under different lattice stretching is considered in the following. Based on Bader charge analysis, the charge transfer between fluorine atoms and 1T-MoS$_2$ was first evaluated. Each F atom gains about 0.49 electrons, meaning that about one hole per unit cell is injected into 1T-MoS$_2$. Therefore, we consider the effect of hole injection and electron injection on the magnetic state of stretched 1T-MoS$_2$, respectively, as summarized in Table S1. It is proved that the energy of the FM state of the system is lower than that of the AFM state when the hole is injected, while the system is still in the AFM ground state when the electron is injected. Figure 3d displays a theoretical model to illustrate the differences caused by different types of charge doping. The arrangement of *d* orbitals of fluorinated 1T-MoS$_2$ has been explained above, where the occupied states below the Fermi energy level are mainly degenerate $d_{xy}$ and $d_{x^2-y^2}$ bands. When a hole is injected, the $d_{xy}$ and $d_{x^2-y^2}$ bands orbitals are less than half-filled, and the exchange interaction between adjacent Mo atoms allows the spins to be aligned in the

same direction, resulting in the FM ground state of the system. But when an electron is injected, it fills the unoccupied band $d_{z^2}$. The $d_{xy}$, $d_{x^2-y^2}$ and $d_{z^2}$ are all half-filled, where the exchange interaction between adjacent Mo atoms allows the spins antiparallel arrangement, leading to the AFM ground state. In general, achieving ideal ferromagnetism in 1T-MoS$_2$ requires a synergy of strain engineering and charge injection. Here we demonstrate that surface chemical adsorption is an effective and feasible method for introducing such synergistic effect in two-dimensional materials.

Finally, the application of 2D materials requires significant consideration of environmental stability. Many 2D magnetic materials face poor environmental stability.[13,14] This is usually due to the presence of H$_2$O or O$_2$ in the air, which easily causes the 2D material to be oxidized or degraded. Here we tested the adsorption behavior of H$_2$O and O$_2$ molecules on the surface of 2D 1T-MoS$_2$F$_2$ (see Figure S4). The adsorption energy is defined as $E_{ab} = E_{total} - E_{MoX_2F_2} - E_{gas}$, where $E_{total}$ represents the total energy of molecular adsorbed on 1T-MoS$_2$F$_2$, $E_{MoX_2F_2}$ represents the energy of 1T-MoS$_2$F$_2$ in a $4 \times 4 \times 1$ supercell, and $E_{gas}$ represents the energy of H$_2$O or O$_2$ molecular. Considering the case where only one molecule is adsorbed on the surface, $E_{ab}$ of H$_2$O and O$_2$ are -110 meV and -70 meV, respectively, and the nearest distances between the absorbed molecules and the surface are more than 2.3 Å, indicating that both H$_2$O and O$_2$ molecular are physically adsorbed on 1T-MoS$_2$F$_2$. The fluorinated MoS$_2$ thus has relatively ideal environmental stability.

We also tested chlorinated and brominated 1T-MoS$_2$. Similarly, chlorinated and brominated MoS$_2$ are also more stable in the 1T phase rather than the 2H phase and have no imaginary frequency in the phonon spectrum (see Figure S5), indicating that 2D 1T-MoS$_2$Cl$_2$ and 1T-MoS$_2$Br$_2$ are thermodynamically stable. The electronic band structures are similar to that of 1T-MoS$_2$F$_2$, both of which are half-metallic (see Figure 4a,c), showing promising applications in future magnetoelectronic devices. Figure 4b, d display the simulated magnetic moment and specific heat with respect to temperature for 1T-MoS$_2$Cl$_2$ and 1T-MoS$_2$Br$_2$, respectively. The T$_C$ of 2D 1T-MoS$_2$Cl$_2$ is 513 K, and the T$_C$ of 2D 1T-MoS$_2$Br$_2$ is 615 K. In addition to halogenated MoS$_2$ monolayers, the

2D fluorinated MoSe$_2$ monolayer can also be converted to ferromagnetism with a T$_C$ of 564 K (Figure S6).

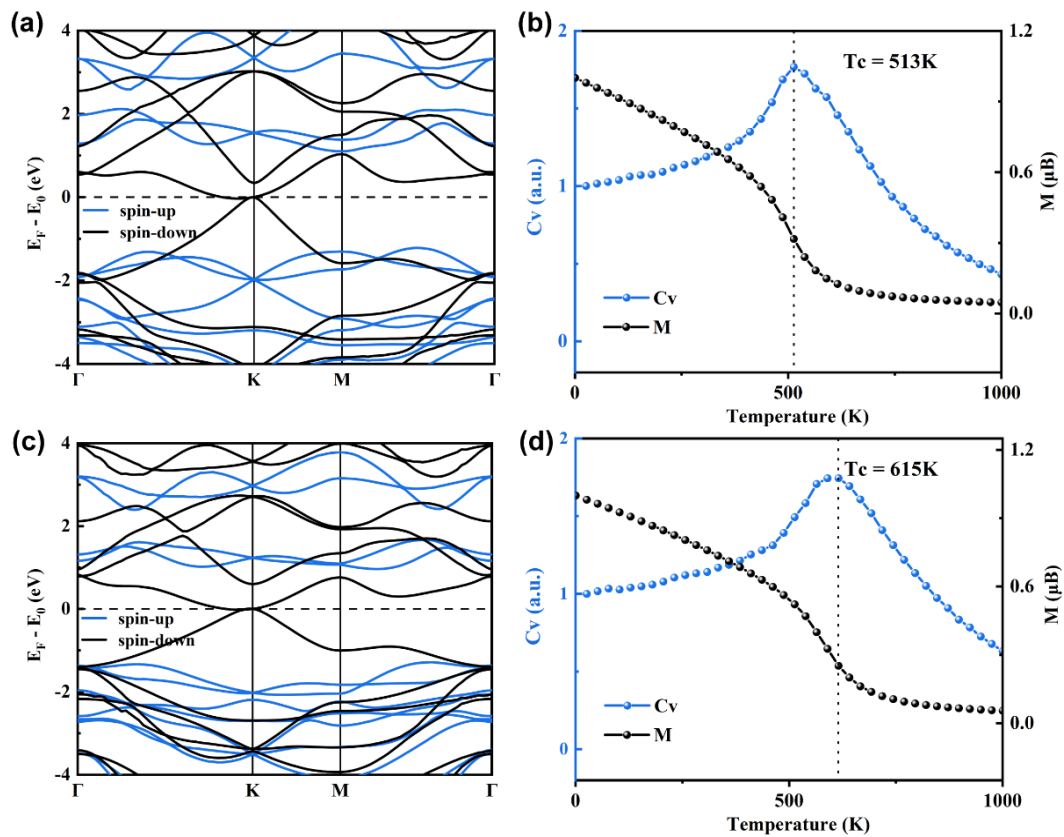

**Figure 4.** (a) Spin-polarized band structure of 1T-MoS$_2$Cl$_2$ monolayer. (b) The simulated magnetic moment (M) and specific heat (C$_v$) with respect to temperature for 1T-MoS$_2$Cl$_2$ monolayer. (c) Spin-polarized band structure of 1T-MoS$_2$Br$_2$ monolayer. (d) The simulated magnetic moment (M) and specific heat (C$_v$) with respect to temperature for 1T-MoS$_2$Br$_2$ monolayer.

**Conclusions**

In conclusion, we demonstrate a promising way to achieve a room temperature 2D magnets basing on non-magnetic MoS$_2$ by the synergistic effect of strain engineering and charge injection. The combined implementation of these two processes is based on the chemical adsorption of halogen atoms (e.g. F, Cl, and Br) on the MoS$_2$'s surface. Unlike primary MoS$_2$ with H-phase, the halogenated MoS$_2$ with T-phase are thermodynamically stable. The adsorbed halogen atoms on the surface leads to lattice superstretching and hole injection, resulting in MoS$_2$ monolayer exhibiting half-metallic properties. The gapless band structure of one spin channel may exhibit interesting topological behaviors, enriching the application of halogenated $MoS_2$ monolayer in the field of low-dimensional magnetoelectronics devices. The Curie temperatures of halogenated monolayer (T-$MoS_2F_2$, T-$MoS_2Cl_2$ and T-$MoS_2Br_2$) are 538 K, 513 K and 615 K, respectively, which are much higher than the room temperature. Lastly, the large magnetic anisotropy energy and good environmental stability make halogenated $MoS_2$ show great practical promising.

**Supporting Information**

The optimized structure and phonon spectrum of $MoS_2F_2$ with H phase; the electronic band structures of 1T-$MoS_2$ under different lattice stretching; two magnetic configurations of 1T-$MoS_2F_2$ monolayer; the optimized structures of $H_2O$ and $O_2$ on the 1T-$MoS_2F_2$'s surface; the phonon spectrum of chlorinated and brominated $MoS_2$.

**Corresponding Author**

* Corresponding Email Address: daoxiong@hainanu.edu.cn; xlli@njnu.edu.cn

**Notes**

The authors declare no competing financial interest.

**Acknowledgment**

This work is supported by the National Natural Science Foundation of China (21803031, 12074190, 22073087), the Postdoctoral Science Foundation of China (No.2018M64043), the Hainan Provincial Natural Science Foundation of China (222MS006), National Natural Science Foundation for Distinguished Young Scholars (22225301), National Key R&D Program of China (2018YFA0208603), CAS project for Young Scientists in Basic Research (YSBR-004), the Start-up Research Foundation of Hainan University (KYQD(ZR)-21125).

# Supporting information

## Turning non-magnetic two-dimensional transition metal sulfides into room temperature magnets by surface fluorination

Jing Wu,[ac] Ruyi Guo,[a] Daoxiong Wu,[b]* Xiuling Li,[ad]* and Xiaojun Wu[e]

[a] School of Physics and Technology, Nanjing Normal University, Nanjing 210023, China

[b] School of Marine Science and Engineering, Hainan Provincial Key Lab of Fine Chemistry, School of Chemical Engineering and Technology, Hainan University, Haikou, 570228, China

[c] Yancheng Kangju Road Junior Middle School, Yancheng, 224000, China

[d] National Laboratory of Solid State Microstructures, Nanjing University, Nanjing 210093, China

[e] Hefei National Research Center for Physical Sciences at the Microscale, Collaborative Innovation Center of Chemistry for Energy Materials, CAS Key Laboratory of Materials for Energy Conversion, CAS Center for Excellence in Nanoscience, and School of Chemistry and Materials Sciences, University of Science and Technology of China, Hefei, Anhui 230026, China

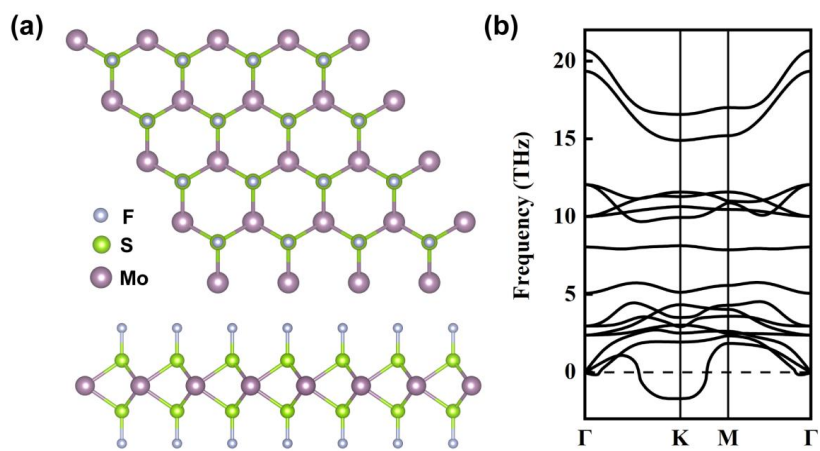

**Figure S1.** (a) Two-dimensional crystal structure and (b) The phonon spectrum of H-phase $MoS_2F_2$.

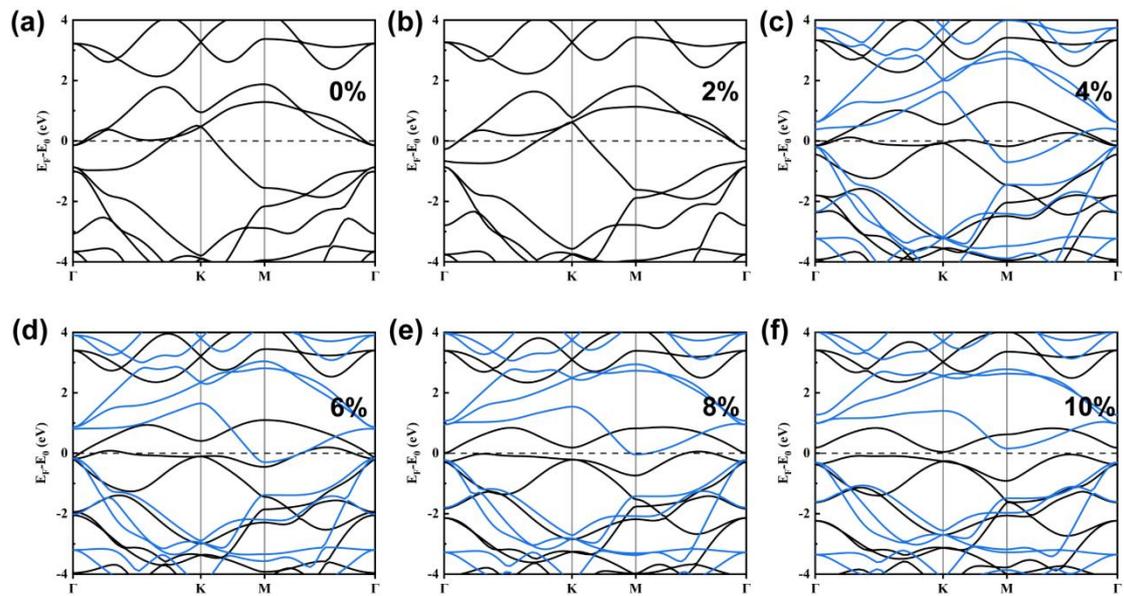

**Figure S2.** The electronic band structures of 1T-MoS$_2$ under different lattice stretching.

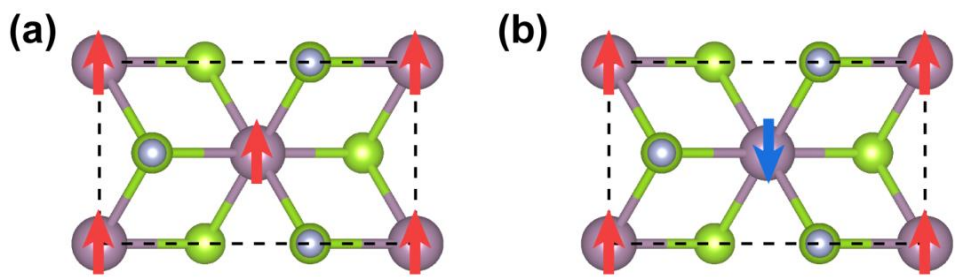

**Figure S3.** Two magnetic configurations of 1T-Mo$_S$2F$_2$ monolayer for (a) FM and (b) AFM, respectively. The "up" and "down" arrows denote the local magnetic moment on Mo atoms.

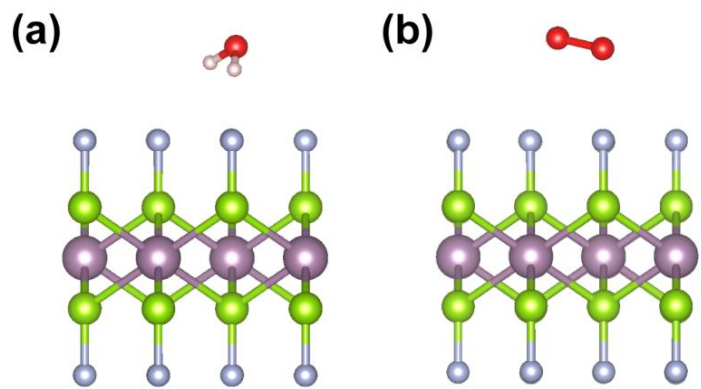

**Figure S4.** The optimized structures of (a) H$_2$O and (b) O$_2$ adsorbed on the 3 × 3 × 1 supercell of 1T-MoS$_2$F$_2$, respectively.

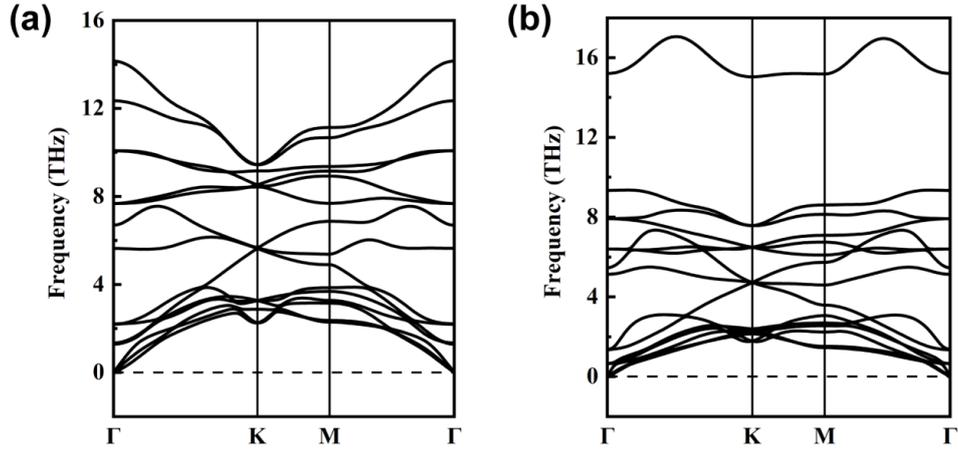

**Figure S5.** The phonon spectrum of (a) chlorinated MoS$_2$ in 1T-phase and (b) brominated MoS$_2$ in 1T-phase, respectively.

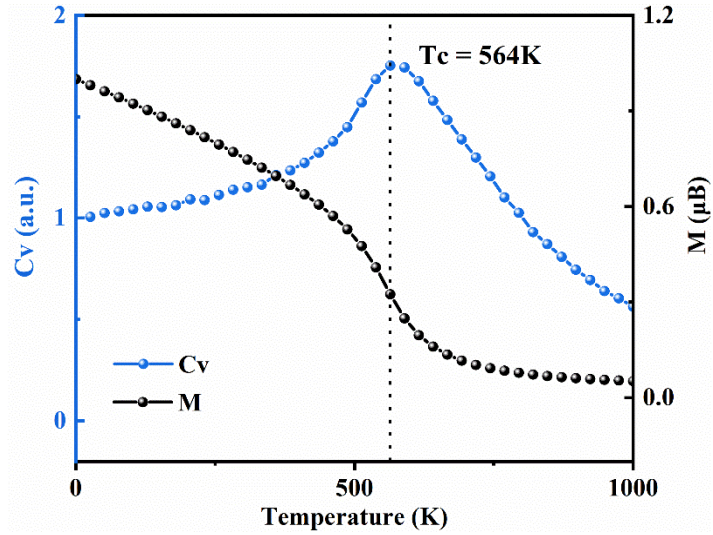

**Figure S6.** The simulated magnetic moment (M) and specific heat ($C_v$) with respect to temperature for 1T- MoSe$_2$F$_2$ monolayer based on Monte Carlo simulations.

**Table S1.** The energy of the FM and AFM states of stretched 1T-MoS$_2$ when the charge and hole are injected respectively.

|  | hole injection | | | | electron injection | | | |
| --- | --- | --- | --- | --- | --- | --- | --- | --- |
|  | 4% | 8% | 12% | 16% | 4% | 8% | 12% | 16% |
| $E_{FM}$/eV | -18.887 | -18.517 | -18.014 | -17.395 | -35.552 | -36.233 | -36.915 | -36.773 |
| $E_{AFM}$/eV | -18.874 | -18.442 | -17.891 | -17.289 | -36.208 | -36.697 | -37.165 | -36.973 |
| $\Delta E$/eV | -0.013 | -0.075 | -0.123 | -0.171 | 0.656 | 0.464 | 0.250 | 0.200 |